\documentclass[preprint,aps,amsmath,showpacs,showkeys,nofootinbib, 
superscriptaddress]{revtex4} 
\usepackage[dvips]{graphicx} 
 
\begin{document} 

\title{\textbf{Coupled channel Faddeev calculations of a $\bar KN\pi$ 
quasibound state}} 

\author{A.~Gal} 
\email{avragal@vms.huji.ac.il} 
\affiliation{Racah Institute of Physics, The Hebrew University, 
Jerusalem 91904, Israel} 

\author{H.~Garcilazo} 
\email{humberto@esfm.ipn.mx} 
\affiliation{Escuela Superior de F\' \i sica y Matem\'aticas \\ 
Instituto Polit\'ecnico Nacional, Edificio 9, 
07738 M\'exico D.F., Mexico} 

\date{\today} 

\begin{abstract} 
The $\bar KN\pi$ system is studied using separable interactions fitted to 
data available on the $s$-wave $\bar KN$--$\pi Y$ subsystem and the $p$-wave 
$\pi N$, $\pi Y$, $\pi \pi$ and $\pi \bar K$ subsystems. Three-body 
$\bar KN\pi$--$\pi Y \pi$ coupled channel Faddeev equations with relativistic 
kinematics are solved in search for poles in the complex energy plane. A 
$\bar KN\pi$ quasibound pole with quantum numbers $I(J^P)=1({\frac{3}{2}}^-)$ 
is found near and below the $\bar KN\pi$ threshold, its precise location 
depending sensitively on the poorly known shape of the $p$-wave $\pi Y$ 
interaction. This $\bar K N \pi$ quasibound state suggests the existence 
of a $D_{13}$ $\Sigma$ resonance with width about 60 MeV near threshold 
($M \approx 1570$ MeV), excluding meson absorption contributions. 
\end{abstract} 



\maketitle 

\newpage

\section{Introduction} 
\label{sec:intro} 

Meson assisted dibaryons are three-body systems consisting of two unbound 
baryons plus a $p$-wave pion \cite{gal11}. For strangeness ${\cal S}=-1$, 
a prototype $I(J^P)=\frac{3}{2}(2^+)$ $Y N \pi$ quasibound state 
($Y\equiv$~hyperon), driven by the $p$-wave resonances $\Delta(1232)$ and 
$\Sigma(1385)$, was studied recently \cite{gg08,gg10}. Replacing a baryon in 
this system by a meson, one obtains a two-meson assisted baryon. For example, 
staying within ${\cal S}=-1$ and substituting $\bar K$ meson for hyperon $Y$, 
or $\pi$ meson for nucleon $N$, a coupled channel $\bar KN\pi$--$Y\pi\pi$ 
two-meson assisted baryon resonance is obtained. In the present paper we study 
this three body system by solving coupled channel, kinematically relativistic 
Faddeev equations. Considering the three-body $\bar KN\pi$ channel, its 
$\bar KN$ subsystem is dominated by the $I(J^P)=0({\frac{1}{2}}^-)$ channel 
in which the $\bar KN -\pi\Sigma$ coupled channel $s$-wave interaction is 
already sufficiently strong to bind, resulting in the $s$-wave $\Lambda$(1405) 
resonance. The $\pi N$ subsystem is dominated by the $I(J^P)=\frac{3}{2}({
\frac{3}{2}}^+)$ channel, resulting in the $p$-wave $\Delta(1232)$ resonance, 
and the $\pi\bar K$ subsystem is dominated by the $I(J^P)=\frac{1}{2}(1^-)$ 
channel, resulting in the ${\bar K}^*(892)$ resonance. In addition, since the 
$\bar KN - \pi Y$ coupling connects $\bar KN\pi$ to $\pi Y\pi$, one must also 
consider the $\pi Y$ subsystem in the $I(J^P)=1({\frac{3}{2}}^+)$ channel, 
resulting in the $p$-wave $\Sigma(1385)$ resonance, and the $\pi\pi$ subsystem 
in the $I(J^P)=1(1^-)$ channel, resulting in the $p$-wave $\rho(770)$ 
resonance. It is straightforward to see that the only three-body configuration 
in which these subsystem quantum number specifications can be accommodated 
is $I(J^P)=1({\frac{3}{2}}^-)$. The value of spin $J^P={\frac{3}{2}}^-$ 
is maximal for a $p$-wave pion and $s$-wave nucleon and $\bar K$ meson, 
thus ensuring that each one of the two-body channels has precisely the 
spin at which it resonates. The value of isospin $I=1$ is not maximal, 
which means that nonresonating two-body channels will also contribute 
to the binding energy balance of $I(J^P)=1({\frac{3}{2}}^-)$ $\bar KN\pi$. 
The corresponding interactions are disregarded with respect to those in the 
resonating channels in this exploratory calculation. Other allowed values of 
isospin, $I=0,2$, stand no chance of producing three-body binding because 
only one of the three possible two-body resonating channels can contribute to 
the $I=0,2$ $\bar KN\pi$ states. Thus, the choice $I(J^P)=1({\frac{3}{2}}^-)$ 
is unique in searching for quasibound configurations of $\bar KN\pi$. It is 
worth noting that the $\bar KN\pi$ threshold around 1570 MeV is within a close 
reach of the one-star $I(J^P)=1({\frac{3}{2}}^-)$ $\Sigma(1580)$ `exotic' 
resonance \cite{PDG}. This apparent connection is discussed towards the 
conclusion of the present paper. 

Other studies of two-meson assisted baryonic resonances focused entirely 
on $s$-wave two-body interactions, for nonstrange resonances \cite{KTO08,
JKE08,TKO09,MTJ10} as well as for strangeness ${\cal S}=-1$ \cite{TKO08} 
and ${\cal S}=-2$ \cite{KEJ08}. For ${\cal S}=-1$, in particular, the 
$I(J^P)=0,1({\frac{1}{2}}^+)$ $\bar KN\pi$ configurations were studied, 
with $\Lambda$ and $\Sigma$ resonance candidates in the mass range 
$M \sim 1.6 - 1.8$ GeV. It is remarkable that with meson-baryon $p$-wave 
interactions, as studied here, a lower mass value can be reached which, 
furthermore, corresponds to a quasibound state rather than resonance. 
Finally, we mention the $KN\pi$ Faddeev calculation in Ref.~\cite{KTO09}, 
again with $s$-wave two-body interactions, searching for a ${\frac{1}{2}}^+$ 
exotic ${\cal S}=+1$ $\Theta^+$ pentaquark. We have verified by solving the 
appropriate Faddeev equations with a $p$-wave pion that no ${\frac{3}{2}}^-$ 
$KN\pi$ quasibound state candidate exists in the relevant energy range for 
a ${\cal S}=+1$ $\Theta^+$ pentaquark. 

The paper is organized as follows. In Sec.~\ref{sec:sepint} we construct 
two-body separable interactions for the resonating channels discussed above. 
The corresponding $t$ matrices serve as input to the set of three-body 
coupled Faddeev equations which are constructed in Sec.~\ref{sec:fad}. 
We employ a straightforward generalization of the nonrelativistic Faddeev 
equations, incorporating relativistic kinematics to account in a minimal way 
for the light pion in the $\bar KN\pi$--$\pi Y\pi$ coupled three-body systems. 
Solving these Faddeev equations we find a $\bar KN\pi$ quasibound pole which 
is listed for several allowed parametrizations of the $\pi Y$ two-body data 
and is discussed in Sec.~\ref{sec:res}. Our calculations, suggesting a 
$D_{13}$ $\Sigma$ resonance near the $\bar KN\pi$ threshold ($M \approx 1570$ 
MeV), are summarized in the last Sec.~\ref{sec:summ}.

\section{Separable two-body interactions input} 
\label{sec:sepint} 

Data available on the $\bar KN$--$\pi Y$, $\pi N$, $\pi\bar K$, and $\pi\pi$ 
subsystems were fitted with rank-one energy independent separable potentials, 
as detailed below. While the $p$-wave subsystems in the mass range considered 
in this work are dominated each by a single resonance pole with no other 
nearby threshold to introduce additional significant energy dependence, 
this may not hold for the $s$-wave $\bar KN$--$\pi Y$ subsystem 
where energy dependence affects the number and position of poles 
(see Ref.~\cite{HJ11} for a recent review). However, energy dependent 
potentials are known to cause problems in relativistically formulated 
three-body calculations \cite{ikeda10}. We further remark on energy 
dependence in subsection \ref{sec:inmedium}. For a standard classification 
of the two-body subsystems, we denote $\bar K$ as particle 1, $N$ as particle 
2, $\pi$ as particle 3, and label the two-body $t$-matrices by the spectator 
particle. Thus, $t_1$ is the $\pi N$ $t$-matrix, $t_2$ is the $\bar K\pi$ 
$t$-matrix, and $t_3$ is the $\bar KN$ $t$-matrix. In addition, we introduce 
$t_4$ as the $\pi\pi$ $t$-matrix. 

\subsection{The $\bar KN-\pi\Sigma-\pi\Lambda$ subsystem} 

The $\bar KN$ interaction allows particle conversion to $\pi\Sigma$ and 
$\pi\Lambda$. There are two resonating two-body channels that contribute 
to the $I(J^P)=1({\frac{3}{2}}^-)$ three-body state. One is the 
$I(J^P)=0({\frac{1}{2}}^-)$ channel which results in the $s$-wave 
$\Lambda$(1405) resonance and the other one is the $I(J^P)=1({\frac{3}{2}}^+)$ 
channel, resulting in the $p$-wave $\Sigma(1385)$ resonance. We use the 
Lippmann-Schwinger equation with relativistic kinematics 
\begin{eqnarray} 
t_3^{\alpha\beta}(p,p'';w_0)= && V_3^{\alpha\beta}(p,p'')+\sum_
{\gamma=\bar KN,\pi\Sigma,\pi\Lambda}\int_0^\infty\, {p'}^2dp'
V_3^{\alpha\gamma}(p,p')\frac{1}{w_0-w_\gamma(p')+i\epsilon} \nonumber \\ 
&& \times t_3^{\gamma\beta}(p',p'';w_0);\,\,\,\, 
\alpha,\beta=\bar KN,\pi\Sigma,\pi\Lambda, 
\label{eq1} 
\end{eqnarray} 
where $w_0$ is the invariant mass of the two-body subsystem and 
\begin{equation} 
w_{ab}(p)=\sqrt{m_a^2+p^2}+\sqrt{m_b^2+p^2}. 
\label{eq2} 
\end{equation} 
Using the separable interaction 
\begin{equation} 
V_3^{\alpha\beta}(p,p')= g_\alpha(p)\lambda_3 g_\beta(p'), 
\label{eq3} 
\end{equation} 
the solution of Eq.~(\ref{eq1}) is 
\begin{equation} 
t_3^{\alpha\beta}(p,p';w_0)=g_\alpha(p)\tau_3(w_0)g_\beta(p'), 
\label{eq4} 
\end{equation} 
with 
\begin{equation} 
{\tau_3}^{-1}(w_0)=\frac{1}{\lambda_3}-
\sum_{\alpha=\bar KN,\pi\Sigma,\pi\Lambda}
\int_0^\infty\,p^2dp\frac{g_\alpha^2(p)}{w_0-w_\alpha(p)+i\epsilon}. 
\label{eq5} 
\end{equation} 

\subsubsection{$s$ waves} 

\begin{figure}[hbt] 
\includegraphics[scale=0.4]{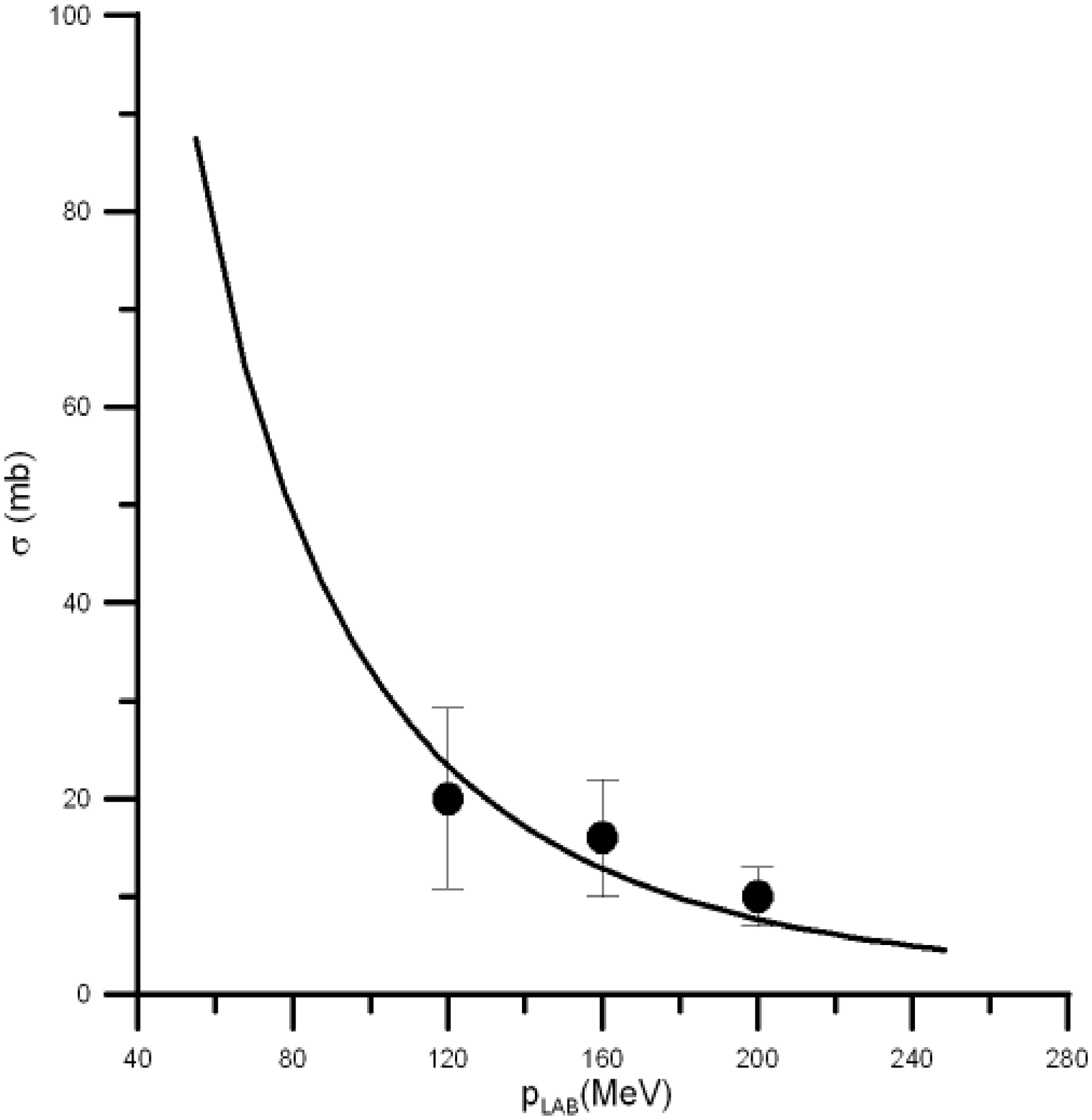} 
\caption{Low energy $K^- p \to \pi^0 \Sigma^0$ cross sections \cite{kim66}. 
The solid curve results from the $I_3=0$ separable interaction (\ref{eq6}) 
with parameters listed in Table I.} 
\label{fig1} 
\end{figure} 

In the case of the isospin $I_3=0$ $\Lambda(1405)$ $s$-wave resonance, 
the $I_{\pi\Lambda}=1$ $\pi\Lambda$ channel is excluded by isospin 
conservation, so that only the channels $\bar KN$ and $\pi\Sigma$ contribute. 
The corresponding form factors were parametrized by Yamaguchi forms: 
\begin{equation} 
g^{s}_{\bar KN}(p)=\frac{1}{c_{\bar KN}^2+p^2},\,\,\,\,\,\,
g^{s}_{\pi\Sigma}(p)=\frac{1}{c_{\pi\Sigma}^2+p^2},  
\label{eq6} 
\end{equation} 
with $g^{s}_{\pi\Lambda}(p)=0$. It is worth pointing out that a single-channel 
energy independent separable potential of the form (\ref{eq3}) does not 
generate $s$-wave resonances, which becomes possible upon including a second 
channel. Note that even with three channels, Eq.~(\ref{eq3}) is a rank-one 
separable potential since it has only one strength parameter $\lambda_3$. The 
three parameters of our $I_3=0$ $\bar KN$--$\pi\Sigma$ separable interaction 
model, $\lambda^{s}_3$, $c_{\bar KN}$, and $c_{\pi\Sigma}$ are listed in 
Table I. They were adjusted to reproduce the PDG position and width of the 
$\Lambda(1405)$ resonance \cite{PDG} as well as the low-energy cross sections 
shown in Fig.~\ref{fig1} for the pure $I_3=0$ $K^- p \to \pi^0 \Sigma^0$ 
reaction. We did not use the more comprehensive data from $K^- p \to 
\pi^{\pm} \Sigma^{\mp}$ because these reactions require information on the 
$I_3=1$ $\bar KN$--$\pi\Sigma$--$\pi\Lambda$ $s$-wave subsystem which is 
nonresonant and is excluded from the present calculation. Preliminary test 
calculations including the $I_3=1$ $s$-wave channel produced negligible 
effects.               

\begin{table}[hbt] 
\caption{Parameters of the $I_3=0$ $\bar KN-\pi\Sigma$ separable $s$-wave 
interaction.} 
\begin{ruledtabular} 
\begin{tabular}{ccccc} 
& $\lambda^{s}_3({\rm fm}^{-2})$ & $c_{\bar KN}({\rm fm}^{-1})$ & 
$c_{\pi\Sigma}({\rm fm}^{-1})$ &  \\ 
\hline 
& $-$3.3391 & 2.0 & 2.5346 &  \\  
\end{tabular} 
\end{ruledtabular} 
\end{table} 

\subsubsection{$p$ waves} 

In the case of the $I_3=1$ $\Sigma(1385)$ $p$-wave resonance, the three 
channels $\bar KN$, $\pi\Sigma$, and $\pi\Lambda$ are all allowed; however, 
there is some evidence that the $\bar KN$ channel couples very weakly to the 
$\Sigma(1385)$ resonance \cite{kim67,martin81} so that one may consider only 
the $\pi\Lambda$ and $\pi\Sigma$ channels. Since in this case the form factors 
$g_\alpha(p)$ must be of a $p$-wave type, we took them in the form
\begin{equation} 
g^{p}_{\pi\Lambda}(p)=p(1+Ap^2)e^{-p^2/\gamma^2},\,\,\,\,\,\, 
g^{p}_{\pi\Sigma}(p)=Bg^{p}_{\pi\Lambda}(p), 
\label{eq7} 
\end{equation} 
with $g^{p}_{\bar KN}(p)=0$. In this case we know only the position, 
width, and branching ratio for decay of the $\Sigma(1385)$ resonance 
into the $\pi\Lambda$ and $\pi\Sigma$ channels. Thus, we have three 
pieces of data to fit our interaction model which contains four parameters
($\lambda^{p}_3$, $\gamma$, $A$, $B$) so that we can take one of these as 
a free parameter and fit the other three. We show in Table II the parameters
$\lambda^{p}_3$, $\gamma$ and $B$, upon gridding on $A$ between 0 to 0.5. 
These parameters differ from those used in the nonrelativistic calculation 
of Ref.~\cite{gg10}. 

\begin{table}[hbt] 
\caption{Parameters of the $\pi\Sigma-\pi\Lambda$ separable interaction 
(\ref{eq7}) in the $I_3=1$ channel for several values of the parameter 
$A$. Values of the r.m.s. momentum ${<p^2>_g}^{\frac{1}{2}}$ (in fm$^{-1}$) 
of the form factor $g^{p}_{\pi Y}(p)$, the r.m.s. distance 
${<r^2>_{\tilde g}}^{\frac{1}{2}}$ and the zero $r_0$ (both in fm) of the 
Fourier transform ${\tilde g}^{p}_{\pi Y}(r)$ are also listed.} 
\begin{ruledtabular} 
\begin{tabular}{ccccccc} 
$A({\rm fm}^2)$ & $\lambda^{p}_3({\rm fm}^{4})$ & $\gamma({\rm fm}^{-1})$ & 
$B$ & ${<p^2>_g}^{\frac{1}{2}}$ & ${<r^2>_{\tilde g}}^{\frac{1}{2}}$ & $r_0$ \\
\hline 
0.00 & $-$0.0077258  & 4.9091  & 0.87039 & 6.94 & 0.58 &  --  \\ 
0.05 & $-$0.0083840  & 3.6156  & 0.87871 & 5.79 & 0.56 & 1.11 \\ 
0.10 & $-$0.0088725  & 3.1595  & 0.89220 & 5.16 & 0.52 & 1.19 \\ 
0.15 & $-$0.0091172  & 2.8951  & 0.90689 & 4.77 & 0.47 & 1.25 \\ 
0.20 & $-$0.0092049  & 2.7147  & 0.92184 & 4.50 & 0.41 & 1.31 \\ 
0.25 & $-$0.0091851  & 2.5810  & 0.93671 & 4.30 & 0.33 & 1.36 \\ 
0.30 & $-$0.0090934  & 2.4765  & 0.95132 & 4.13 & 0.23 & 1.41 \\ 
0.35 & $-$0.0089513  & 2.3919  & 0.96559 & 4.00 &  --  & 1.45 \\ 
0.40 & $-$0.0087763  & 2.3216  & 0.97949 & 3.89 &  --  & 1.48 \\ 
0.45 & $-$0.0085787  & 2.2619  & 0.99298 & 3.80 &  --  & 1.51 \\ 
0.50 & $-$0.0083686  & 2.2105  & 1.00606 & 3.72 &  --  & 1.54 \\ 
\end{tabular} 
\end{ruledtabular} 
\end{table} 

Listed also in the table are ${<p^2>}^{\frac{1}{2}}$ values for 
$g^{p}_{\pi Y}(p)$, plus ${<r^2>}^{\frac{1}{2}}$ and zeros $r_0$ for its 
Fourier transform ${\tilde g}^{p}_{\pi Y}(r)$. These form-factor sizes will 
be discussed and compared in subsection \ref{subsec:size} with those for the 
$p$-wave form factors derived below for other subsystems.

\subsection{The $\pi N$ subsystem} 

\begin{figure}[hbt]  
\includegraphics[scale=0.4]{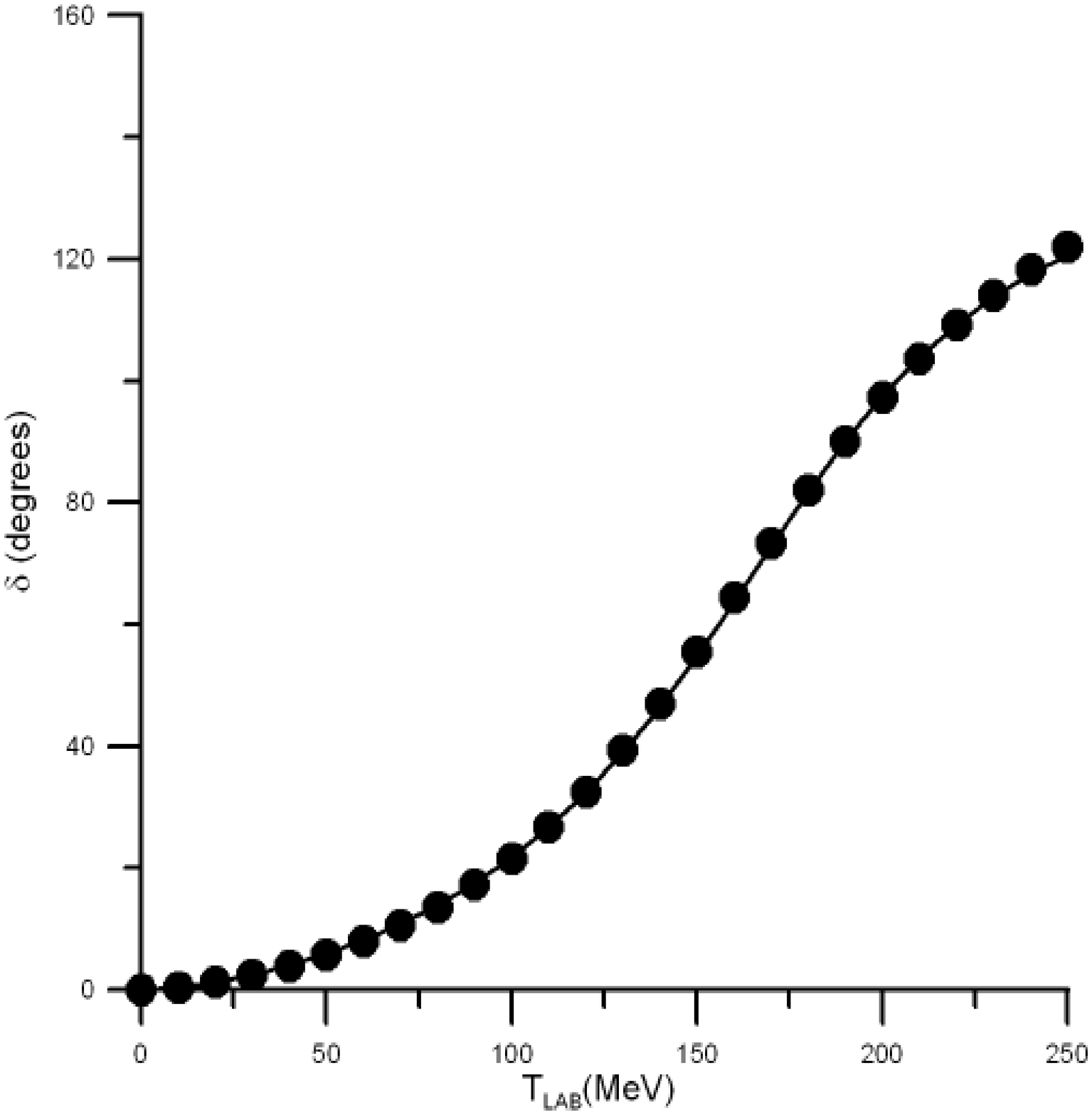} 
\caption{The $\pi N$ $P_{33}$ phase shift across the $\Delta(1232)$ resonance. 
The solid curve is obtained by using the $\pi N$ form factor parameters, 
Eq.~(\ref{eq11}), listed in Table III. The circles are from Ref.~\cite{ARNDT} 
with errors (suppressed in the figure) that are less than the thickness of the 
line.} 
\label{fig2} 
\end{figure} 

The $\pi N$ $p$-wave interaction is dominated by the $P_{33}$ channel through 
the $\Delta(1232)$ resonance and we use for it, as well as for 
the remaining two-body subsystems, a rank-one separable potential 
\begin{equation} 
V_1(p,p')=g_1(p)\lambda_1 g_1(p'), 
\label{eq8} 
\end{equation} 
the $t$-matrix of which is 
\begin{equation} 
t_1(p,p';w_0)=g_1(p)\tau_1(w_0)g_1(p'), 
\label{eq9}  
\end{equation} 
with 
\begin{equation} 
{\tau_1}^{-1}(w_0)=\frac{1}{\lambda_1}-\int_0^\infty\, p^2dp\frac{g_1^2(p)}
{w_0-w_{\pi N}(p)+i\epsilon}. 
\label{eq10} 
\end{equation} 
The $\pi N$ form factor, as well as for the subsequent two-body subsystems, 
is chosen as 
\begin{equation} 
g_1(p)=p[e^{-p^2/\beta^2}+Cp^2e^{-p^2/\alpha^2}]. 
\label{eq11} 
\end{equation} 
We show in Fig.~\ref{fig2} the fit to the $P_{33}$ phase shift of 
Ref.~\cite{ARNDT} obtained with the set of $\pi N$ parameters given 
in the first line of Table III.

\subsection{The $\pi\bar K$ subsystem} 

\begin{figure}[hbt]  
\includegraphics[scale=0.4]{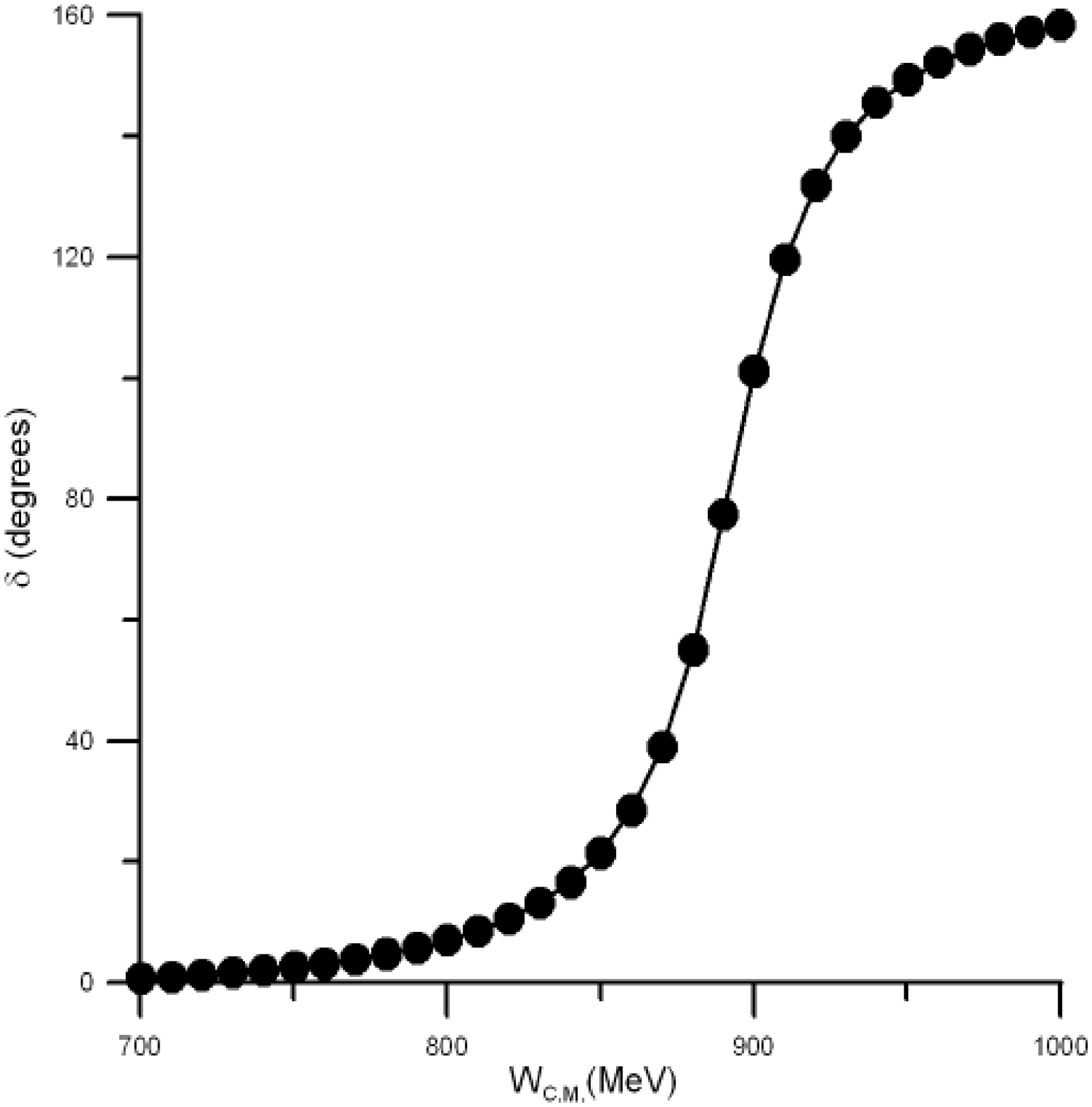} 
\caption{The $I_2=\frac{1}{2},J_2=1$ $p$-wave $\pi \bar K$ phase shift across 
the ${\bar K}^*(892)$ resonance. The solid curve is obtained by using the 
$\pi \bar K$ form factor parameters, Eq.~(\ref{eq11}), listed in Table III. 
The circles are from Ref.~\cite{BEJ10} with errors (suppressed in the figure) 
that are less than the thickness of the line.} 
\label{fig3} 
\end{figure} 

The separable interaction $V_2(p,p')=g_2(p)\lambda_2 g_2(p')$ corresponding 
to the $\pi\bar K$ subsystem is chosen such that the $\pi\bar K$ form factor 
$g_2(p)$ is functionally identical with $g_1(p)$, Eq.~(\ref{eq11}), for 
$\pi N$. To fix the parameters of $g_2(p)$, we used the phase shift by 
Boito {\it et al.} fitted to $\tau \to K \pi \nu_{\tau}$ and $K_{\ell_3}$ 
decays \cite{BEJ10}. These parameters are listed in the second line of Table 
III, and the agreement with this phase shift is shown in Fig.~\ref{fig3}.

\subsection{The $\pi\pi$ subsystem} 

\begin{figure}[hbt]  
\includegraphics[scale=0.4]{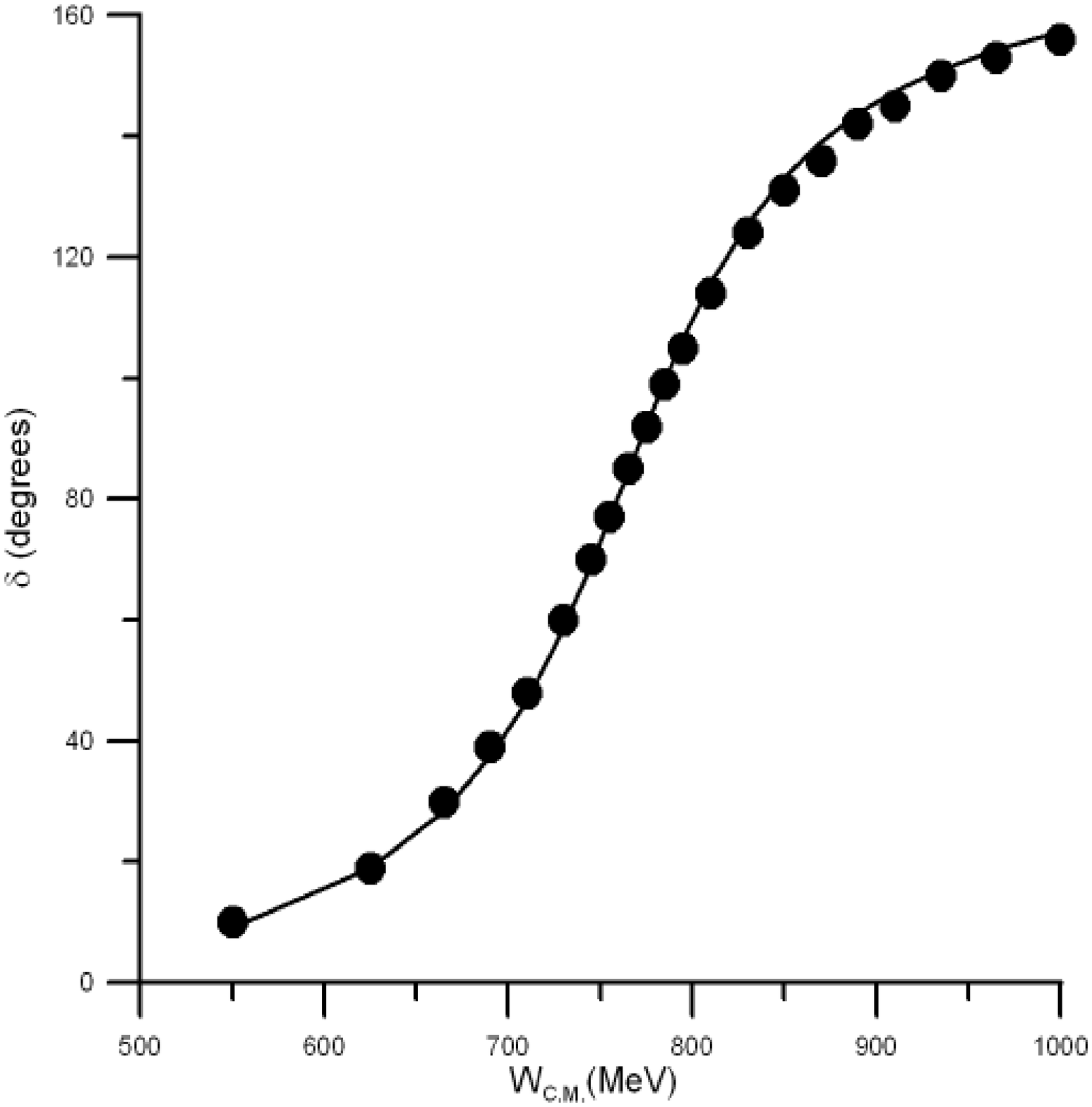} 
\caption{The $I_4=1,J_4=1$ $p$-wave $\pi \pi$ phase shift across the 
$\rho(770)$ resonance. The solid curve is obtained by using the $\pi \pi$ form 
factor parameters, Eq.~(\ref{eq11}), listed in Table III. The circles are from 
Ref.~\cite{PROTO} with slight errors suppressed in the figure.} 
\label{fig4} 
\end{figure} 

Since this subsystem is reached only following ${\bar K}N \to \pi Y$ 
conversion, we denote its $t$ matrix by $t_4$. The separable interaction 
$V_4(p,p')=g_4(p)\lambda_4 g_4(p')$ is given again by a form factor $g_4(p)$ 
which is functionally identical with $g_1(p)$, Eq.~(\ref{eq11}), for $\pi N$. 
We fitted the $\pi\pi$ $I=J=1$ $p$-wave phase shift obtained in 
Ref.~\cite{PROTO}, as shown in Fig.~\ref{fig4}, using the set of parameters 
given in the last line of Table III. 

\begin{table}[hbt]  
\caption{Parameters of the $\pi N$, $\pi\bar K$, and $\pi\pi$ separable 
$p$-wave interactions, with form factors $g(p)$ given by Eq.~(\ref{eq11}). 
Values of the r.m.s. momentum ${<p^2>_g}^{\frac{1}{2}}$ (in fm$^{-1}$), 
the r.m.s. distance ${<r^2>_{{\tilde g}}}^{\frac{1}{2}}$ of the Fourier 
transform ${\tilde g}(r)$ and its zero $r_0$ (both in fm) are also listed.} 
\begin{ruledtabular} 
\begin{tabular}{cccccccc} 
 subsystem & $\lambda({\rm fm}^4)$ & $\alpha({\rm fm}^{-1})$ & 
$\beta({\rm fm}^{-1})$ & $C({\rm fm}^2)$ & 
${<p^2>}^{\frac{1}{2}}$ & ${<r^2>}^{\frac{1}{2}}$ & $r_0$ \\ 
\hline 
$\pi N$  & $-$0.075869 & 2.3668 & 1.04 & 0.23 & 4.07 & 1.47 & 1.36 \\ 
$\pi\bar K$  & $-$0.0037111 & 4.3057 & 1.703 & 0.122 & 7.46 & 0.93 & 0.74 \\ 
$\pi\pi$     & $-$0.0078958 & 5.6646 & 1.89 & 0.03  &  9.81 & 0.88 & 0.56 \\ 
\end{tabular} 
\end{ruledtabular} 
\end{table}

\subsection{Form factor sizes} 
\label{subsec:size} 

The form factors $g(p)$, Eq.~(\ref{eq11}), with parameters fitted to 
phase shift analyses of the corresponding $p$-wave resonances as listed 
in Table III, all have r.m.s. values ${<p^2>}^{\frac{1}{2}}$ larger than 
4~fm$^{-1}$, suggesting naively spatial extensions of less than 0.25~fm. 
If this were true, the usefulness of treating the $\bar K N\pi$ system 
in terms of hadronic degrees of freedom would have become questionable. 
However, the uncertainty principle places only a lower bound on the 
spatial extension and indeed the ${<r^2>}^{\frac{1}{2}}$ values listed 
in the table are considerably larger than 0.25~fm, thus qualifying for 
hadronic attributes. These r.m.s. radii were calculated for ${\tilde g}(r)$, 
where ${\tilde g}(\vec r\:)={\hat r}{\tilde g}(r)$ is the Fourier transform 
of the $p$-wave form factor $g(\vec p\:)={\hat p}g(p)$. Up to a constant, 
\begin{equation} 
{\tilde g}(r)=\int j_1(pr)g(p)p^2dp, 
\label{eq11a} 
\end{equation} 
with $j_1$ the spherical Bessel function for $\ell = 1$. Unlike $g(p)$, 
${\tilde g}(r)$ is not positive definite, it flips from positive below 
$r_0$ to negative above $r_0$. This may result in negative values 
of $<r^2>$, as observed in Table II for some of the $\pi Y$ form factors, 
and in grossly underestimated sizes for those form factors for which $<r^2>$ 
is still positive. For this reason, we prefer to use $r_0$ as an alternative 
spatial size parameter. The values assumed by $r_0$ in Table III, although 
smaller than the corresponding ${<r^2>}^{\frac{1}{2}}$ values, still qualify 
for being considered as hadronic sizes. It is reassuring that the values of 
$r_0$ for $\pi Y$ in Table II are all larger than 1 fm. A crude way to relate 
the expected range of values of $r_0$ for $\pi Y$ to the value of $r_0$ for 
$\pi N$ is to note that the energy excitation from $\Lambda$ to $\Sigma(1385)$ 
is somewhat less than from $N$ to $\Delta(1232)$, and therefore $r_0$ for 
$\pi Y$ should be a bit larger than $r_0$ for $\pi N$, judging also by the 
systematics of $\pi \bar K$ with respect to $\pi\pi$ in Table III. 
We therefore estimate that $r_0$ for $\pi Y$ is roughly up to 0.15 fm larger 
than for $\pi N$, i.e., in the range 1.36--1.51 fm. This argument suggests 
that among the $\pi Y$ form factors listed in Table II, only those in the 
range $A=0.25$ to $A=0.45$ fm$^2$ are consistent with the form-factor 
phenomenology used for the other subsystems.

\section{Three-body equations} 
\label{sec:fad} 

\subsection{Two-body $t$-matrix in a three-body system} 
\label{sec:inmedium} 

To embed the two-body subsystems discussed in Sec.~\ref{sec:sepint} 
into the three-body system, Eq.~(\ref{eq10}) needs to be generalized to 
\begin{equation} 
{\tau_1}^{-1}(W_0;q)=\frac{1}{\lambda_1}-\int_0^\infty\, p^2dp\frac{g_1^2(p)}
{W_0-\sqrt{q^2+m_{\bar K}^2}-\sqrt{q^2+w_{\pi N}^2(p)}+i\epsilon}, 
\label{eq12} 
\end{equation} 
where $W_0$ is the invariant mass of the three-body system and $m_{\bar K}$ 
and $q$ are the mass and momentum of the spectator particle in the three-body 
c.m. frame. Similar expressions apply to the other subsystems. 
In the case of the amplitude $t_4$ the mass of the spectator particle can 
be either $m_\Sigma$ or $m_\Lambda$ so that this amplitude is of the form 
$\tau_4^Y(W_0;q)$. We note that in the three-body c.m. frame, in which 
the Faddeev equations are formulated and solved, the two-body `in medium' 
isobar propagators input $\tau(W_0;q)$ are energy dependent even though the 
underlying two-body separable potentials are energy independent. The $\tau$'s 
depend also on the spectator momentum $q$, reducing to the two-body momentum 
independent $t$'s of Eq.~(\ref{eq10}) for $q=0$.

\subsection{Three-body Faddeev equations} 
\label{sec:3body} 

Allowing for particle conversion, there are two possible three-body 
states $\bar KN\pi$ and $\pi Y\pi$ which we will refer as $a$ and $b$, 
respectively. Therefore, the Green's function for three free particles 
is a $2\times 2$ matrix of the form 
\begin{equation} 
G_0=\begin{pmatrix}G_a & 0 \cr 
              0  & G_b \cr\end{pmatrix}= 
\begin{pmatrix}G_a & 0 \cr 
           0  & 0 \cr\end{pmatrix}+ 
\begin{pmatrix}0 & 0 \cr 
         0  & G_b \cr\end{pmatrix}\equiv 
              G_0^a+G_0^b, 
\label{eq13} 
\end{equation} 
with 
\begin{equation} 
G_a=\frac{1}{W_0-\sqrt{m_{\bar K}^2+q_1^2}-\sqrt{m_N^2+q_2^2}
-\sqrt{m_\pi^2+q_3^2}+i\epsilon}, 
\label{eq14} 
\end{equation} 
\begin{equation} 
G_b=\frac{1}{W_0-\sqrt{m_\pi^2+q_1^2}-\sqrt{m_Y^2+q_2^2}
-\sqrt{m_\pi^2+q_3^2}+i\epsilon}. 
\label{eq15} 
\end{equation} 
Since particle conversion is effected only through the amplitude $t_3$, 
the three-body Faddeev equations take the form 
\begin{eqnarray} 
T_1 &= & t_1G_0^aT_2+t_1G_0^aT_3, 
\label{eq16} \\
T_2 &= & t_2G_0^aT_1+t_2G_0^aT_3, 
\label{eq17} \\
T_3 &= & t_3G_0^aT_1+t_3G_0^aT_2+t_3G_0^bT_3+2t_3G_0^bT_4, 
\label{eq18} \\
T_4 &= & t_4G_0^bT_3. 
\label{eq19} 
\end{eqnarray} 
Here, $T_i$ is that part of the three-body amplitude where in the last stage 
particle $i$ is spectator while particles $j$ and $k$ interact. Particle 
conversion is generated exclusively through the two-body amplitude $t_3$ in 
the last two terms on the r.h.s. of Eq.~(\ref{eq18}). $T_4$ represents 
that part of the three-body amplitude where in the last stage the hyperon is 
spectator while the two pions interact, so that $T_4$ couples to $T_3$ in 
Eq.~(\ref{eq18}) through pion exchange and the factor 2 arises because 
any one of the two pions may be exchanged. The amplitude $T_3$ can couple to 
itself as a consequence of hyperon exchange. 

The two-body $t$-matrices constructed in the previous section 
can be written in the space $\begin{pmatrix}a \cr b \cr\end{pmatrix}$ 
as $2\times 2$ matrices of the form 
\begin{eqnarray} 
t_i &=& |g_i\rangle\begin{pmatrix}\tau_i & 0 \cr 0 & 0 \cr\end{pmatrix}
\langle g_i|;\,\,\, i=1,2, 
\label{eq20} \\
t_3 &=& \begin{pmatrix}|g_{\bar KN}\rangle \cr |g_{\pi Y}\rangle \cr
\end{pmatrix}\tau_3 
\begin{pmatrix}\langle g_{\bar KN}| & \langle g_{\pi Y}| \cr
\end{pmatrix}, 
\label{eq21} \\
t_4 &=& |g_4\rangle\begin{pmatrix}0 & 0 \cr 0 & \tau_4 \cr
\end{pmatrix}\langle g_4|, 
\label{eq22} 
\end{eqnarray} 
so that the Faddeev components are of the form 
\begin{eqnarray} 
T_i &=& |g_i\rangle\begin{pmatrix}X_i \cr 0 \cr\end{pmatrix};\,\,\, i=1,2, 
\label{eq23} \\
T_3 &=& \begin{pmatrix}|g_{\bar KN}\rangle \cr |g_{\pi Y}\rangle \cr
\end{pmatrix}X_3, 
\label{eq24} \\
T_4 &=& |g_4\rangle\begin{pmatrix}0  \cr  X_4 \cr\end{pmatrix}. 
\label{eq25} 
\end{eqnarray} 
Substituting Eqs.~(\ref{eq20})-(\ref{eq25}) into the Faddeev equations 
(\ref{eq16})-(\ref{eq19}), we get the equations for the amplitudes $X_i$ 
\begin{eqnarray} 
X_i &=& \tau_i\langle g_i|G_a|g_{3-i}\rangle X_{3-i} 
      + \tau_i\langle g_i|G_a|g_{\bar KN}\rangle X_3,\,\,\,\,\,\,i=1,2 
\label{eq26} \\
X_3 &=& \tau_3\sum_{j=1}^2\langle g_{\bar KN}|G_a|g_j\rangle X_j \nonumber \\ 
      & + & \tau_3\langle g_{\pi Y}|G_b|g_{\pi Y}\rangle X_3 
      + 2\tau_3\langle g_{\pi Y}|G_b|g_4\rangle X_4, 
\label{eq27} \\
X_4 &= & \tau_4\langle g_4|G_b|g_{\pi Y}\rangle X_3. 
\label{eq28} 
\end{eqnarray} 
These equations take the explicit form 
\begin{eqnarray} 
X_i(q_i) &=& \tau_i(W_0;q_i)\int_0^\infty dq_{3-i} 
K_{i(3-i)}(q_i,q_{3-i})X_{3-i}(q_{3-i}) \nonumber \\ 
&+& \tau_i(W_0;q_i)\sum_{I_3=0,1} \int_0^\infty dq_3 
K_{i3}^{I_3}(q_i,q_3)X_3^{I_3}(q_3),\,\,\,i=1,2, 
\label{eq29} \\
X_3^{I_3}(q_3) &=&\tau_3^{I_3}(W_0;q_3)\sum_{j=1}^2\int_0^\infty dq_j 
K_{3j}^{I_3}(q_i,q_j)X_j(q_j) + \tau_3^{I_3}(W_0;q_3) 
\sum_{I_3^\prime=0,1} \int_0^\infty dq_1 
K_{33}^{I_3I_3^\prime}(q_3,q_1) \nonumber \\  
& \times & X_3^{I_3^\prime}(q_1) 
+ 2\tau_3^{I_3}(W_0;q_3)\sum_{Y=\Sigma,\Lambda} 
\int_0^\infty dq_2 
K_{34}^{I_3Y}(q_3,q_2)X_4^Y(q_2),\,\,\,I_3=0,1, 
\label{eq30} \\
X_4^Y(q_2) & = & \tau_4^Y(W_0;q_2) \sum_{I_3=0,1} \int_0^\infty dq_3 
K_{43}^{YI_3}(q_2,q_3)X_3^{I_3}(q_3);\,\,\,Y=\Sigma,\Lambda,  
\label{eq31} 
\end{eqnarray} 
where the dependence of the amplitudes $X_i$ and the kernels $K_{ij}$ on the 
total energy $W_0$ was suppressed. The amplitudes $X_i$ depend each on its 
spectator momentum $q_i$. These spectator momenta are related by the 
three-body c.m. constraint which is evident upon inspecting the expressions 
for the kernels $K_{ij}$: 
\begin{equation} 
K_{12}(q_1,q_2)=\frac{q_1q_2}{2}\int_{-1}^1 d{\cos}\theta\, 
\frac{g_1(p_1)(\hat p_1\cdot\hat p_2)g_2(p_2)b_{12}^{\frac{3}{2}\frac{1}{2}}} 
{W_0-\sqrt{m_{\bar K}^2+q_1^2}-\sqrt{m_N^2+q_2^2}
-\sqrt{m_\pi^2+(\vec q_1+\vec q_2)^2}+i\epsilon}, 
\label{eq32} 
\end{equation} 
\begin{equation} 
K_{31}^{I_3}(q_3,q_1)=\frac{q_3q_1}{2} \int_{-1}^1 d{\cos}\theta\,\frac
{g_{\bar KN}^{I_3}(p_3)(\hat \kappa_3\cdot\hat p_1)g_1(p_1)
b_{31}^{I_3\frac{3}{2}}}{W_0-\sqrt{m_\pi^2+q_3^2}-\sqrt{m_{\bar K}^2+q_1^2}
-\sqrt{m_N^2+(\vec q_3+\vec q_1)^2}+i\epsilon}, 
\label{eq33} 
\end{equation} 
\begin{equation} 
K_{23}^{I_3}(q_2,q_3)=\frac{q_2q_3}{2} \int_{-1}^1 d{\cos}\theta\,\frac
{g_2(p_2)(\hat p_2\cdot\hat \kappa_3)g_{\bar KN}^{I_3}(p_3)
b_{23}^{\frac{1}{2}I_3}} 
{W_0-\sqrt{m_N^2+q_2^2}-\sqrt{m_\pi^2+q_3^2}
-\sqrt{m_{\bar K}^2+(\vec q_2+\vec q_3)^2}+i\epsilon}, 
\label{eq34} 
\end{equation} 
\begin{equation} 
K_{33}^{I_3I_3^\prime}(q_3,q_1)=\frac{q_3q_1}{2} 
\sum_{Y=\Sigma,\Lambda}\int_{-1}^1 d{\cos}\theta\, 
\frac{g_{\pi Y}^{I_3}(p_3)(\hat \kappa_3\cdot\hat \kappa_1) 
g_{\pi Y}^{I_3^\prime}(p_1)b_{31}^{I_3I_3^\prime}} 
{W_0-\sqrt{m_\pi^2+q_3^2}-\sqrt{m_\pi^2+q_1^2}
-\sqrt{m_Y^2+(\vec q_3+\vec q_1)^2}+i\epsilon}, 
\label{eq35} 
\end{equation} 
\begin{equation} 
K_{43}^{YI_3}(q_2,q_3)=\frac{q_2q_3}{2} \int_{-1}^1 d{\cos}\theta\, 
\frac{g_4(p_2)(\hat p_2\cdot\hat \kappa_3)g_{\pi Y}^{I_3}(p_3)b_{23}^{1I_3}}
{W_0-\sqrt{m_\pi^2+q_2^2}-\sqrt{m_{\pi}^2+q_3^3}
-\sqrt{m_Y^2+(\vec q_2+\vec q_3)^2}+i\epsilon}, 
\label{eq36} 
\end{equation} 
where the dependence on the total energy $W_0$ in the arguments of the kernels 
$K_{ij}$ was again suppressed. In Eqs.~(\ref{eq33})-(\ref{eq36}) 
$\hat \kappa_i=\hat q_i$ if $I_3=0$ and $\hat \kappa_i=\hat p_i$ if $I_3=1$.

The isospin recoupling coefficients are 
\begin{equation} 
b_{ij}^{I_iI_j}=(-)^{I_j+\tau_j-I}\sqrt{(2I_i+1)(2I_j+1)}
W(\tau_j\tau_k I \tau_i;I_iI_j), 
\label{eq37} 
\end{equation} 
with $W$ the Racah coefficient. $I=1$ is the total isospin and $\tau_1$, 
$\tau_2$, $\tau_3$ are the isospins of the three particles. In the first 
three kernels the three particles are $\bar K$, $N$, $\pi$ and in the last 
two kernels they are $\pi$, $Y$, $\pi$. 

From Eqs.~(\ref{eq32})-(\ref{eq36}) one obtains the other necessary 
expressions by using $K_{ji}(q_j,q_i)=K_{ij}(q_i,q_j)$, etc. As for $p_i$, 
the magnitude of the relative three-momentum $\vec p_i$, it is Lorentz 
invariant since it is expressible in terms of the invariant mass of the 
relative momentum four-vector, see Eqs.~(28)--(30) in Ref.~\cite{gg08}. 
The details of the relativistic boost involved in expressing $\vec p_i$ 
in the three-body c.m. system were recorded in Eqs.~(32)--(33) there and 
are adapted below to the present notations. Thus, one can calculate $p_i$, 
$p_j$, ($\hat p_i\cdot\hat p_j$), ($\hat q_i\cdot\hat p_j$), and 
($\hat p_i\cdot\hat q_j$) by using 
\begin{equation} 
\vec p_i=-\vec q_j-a_{ij}\vec q_i~,~~~~\vec p_j=\vec q_i+a_{ji}\vec q_j~, 
\label{eq38} 
\end{equation} 
where $i,j$ is a cyclic pair, ${\cos}\theta=\hat q_i\cdot\hat q_j$, and 
\begin{equation} 
a_{ij}=\frac{W_i^2-q_i^2+m_j^2-m_k^2+2\sqrt{(m_j^2+q_j^2)(W_i^2-q_i^2)}}
{2\sqrt{W_i^2-q_i^2}\left(W_i+\sqrt{W_i^2-q_i^2}\right)}, 
\label{eq39} 
\end{equation} 
\begin{equation} 
a_{ji}=\frac{W_j^2-q_j^2+m_i^2-m_k^2+2\sqrt{(m_i^2+q_i^2)(W_j^2-q_j^2)}}
{2\sqrt{W_j^2-q_j^2}\left(W_j+\sqrt{W_j^2-q_j^2}\right)}, 
\label{eq40} 
\end{equation} 
with 
\begin{equation} 
W_i=\sqrt{m_j^2+q_j^2}+\sqrt{m_k^2+(\vec q_i+\vec q_j)^2}, 
\label{eq41} 
\end{equation} 
\begin{equation} 
W_j=\sqrt{m_i^2+q_i^2}+\sqrt{m_k^2+(\vec q_i+\vec q_j)^2}. 
\label{eq42} 
\end{equation} 
Eqs.~(\ref{eq38})-(\ref{eq42}) correspond to relativistic kinematics 
for three particles on the mass shell. 

In order to find the eigenvalues of the integral equations 
(\ref{eq29})-(\ref{eq31}), integrals were replaced by sums applying numerical 
integration quadrature. In this way the equations become a set of homogeneous 
linear equations which have solutions only if the determinant of the matrix 
of its coefficients (the Fredholm determinant) vanishes at certain (complex) 
energies. We used the standard procedure described in Ref.~\cite{PA84}, 
i.e., we make the contour rotation $q_i \to q_i \exp (-{\rm i}\phi)$ which 
opens some portions of the second Riemann sheet for the variable $W_0$. 
This allows one to look for poles of Eqs.~(\ref{eq29})-(\ref{eq31}) by taking 
$W_0=M-i\Gamma/2$, and calculating the Fredholm determinant to look for its 
zeros.

\section{Results and Discussion} 
\label{sec:res} 

Before reporting the results of a full three-body Faddeev calculation for the 
$\bar KN\pi$--$\pi Y \pi$ coupled channels, we discuss some relevant partial 
calculations. 
\begin{itemize} 
\item Limiting the Faddeev equations to the lower $\pi Y \pi$ three-body 
channel, for various combinations of the input two-body interactions, 
resonance poles about 200 MeV above the $\bar KN\pi$ threshold are obtained. 
The $p$-wave $\pi Y$ and $\pi\pi$ interactions used were those described in 
Sec.~\ref{sec:sepint}, whereas separable interactions corresponding to the 
scattering length and effective radius combinations listed by Ikeda 
{\it et al.} \cite{ikeda11} for the $I_3=0$ $\pi \Sigma$ interaction were 
constructed to simulate the $s$-wave $\pi Y$ interaction. This interaction 
which is used in meson-baryon chiral models is too weak to help bind the 
$\pi Y \pi$ system, much the same as it is too weak in a two-body calculation 
to generate on its own a resonance similar to $\Lambda(1405)$, without 
coupling in the $I_3=0$ $\bar KN$ upper channel interaction. 
\item Limiting the Faddeev equations to the upper $\bar KN\pi$ three-body 
channel, for various combinations of the input two-body interactions, 
resonance poles about 100 MeV above the $\bar KN\pi$ threshold are obtained. 
The $p$-wave $\pi N$ and $\pi\bar K$ interactions used were those described 
in Sec.~\ref{sec:sepint}. In these calculations, the $I_3=0$ $\bar KN$ 
$s$-wave interaction used was sufficiently strong to bind on its own in the 
range of $M_{\bar KN} \sim$~1420--1430 MeV. 
\item Using a complex $I_3=0$ $\bar KN$ $s$-wave interaction in the 
$\bar KN\pi$ three-body Faddeev calculations, to simulate implicitly coupling 
to the lower $\pi Y\pi$ channel, the $\bar KN\pi$ resonance energy decreased 
as low as to about 50 MeV above the $\bar KN\pi$ threshold. Here, the $I_3=0$ 
$\bar KN$ scattering length was fixed at $a_3=-1.70+{\rm i}0.68$ fm \cite
{martin81} and the real part of the fitted separable scattering amplitude 
changed sign at around 1405 MeV, with a quasibound pole in the range 
$M_{\bar KN}\sim$~1415--1425 MeV. 
\end{itemize}  
 
\begin{table}[hbt]  
\caption{Energy eigenvalue of the $\bar KN\pi$ system for the 
several models of the $I=1$ $\pi Y$ $p$-wave interaction given in 
Table II and the models of the $\bar KN$, $\bar K\pi$, $\pi N$, and $\pi\pi$ 
interactions given in Tables I and III.} 
\begin{ruledtabular} 
\begin{tabular}{cccc} 
 & $A({\rm fm}^2)$ 
& $W_0-m_{\bar K}-m_N-m_\pi$ (MeV) \\ 
\hline 
 & 0.00 & $-111.0-{\rm i}0.1$ & \\ 
 & 0.05 & $-67.5-{\rm i}1.4$ & \\ 
 & 0.10 & $-42.1-{\rm i}6.4$ & \\ 
 & 0.15 & $-26.1-{\rm i}13.0$ & \\ 
 & 0.20 & $-15.4-{\rm i}20.6$ & \\ 
 & 0.25 & $-7.6-{\rm i}24.6$ & \\ 
 & 0.30 & $-1.2-{\rm i}29.5$ & \\ 
 & 0.35 & $+3.6-{\rm i}32.4$ & \\ 
 & 0.40 & $+7.8-{\rm i}34.3$ & \\ 
 & 0.45 & $+10.9-{\rm i}35.2$ & \\ 
 & 0.50 & $+12.8-{\rm i}37.4$ & \\ 
\end{tabular} 
\end{ruledtabular} 
\end{table} 

Finally, the full $\bar KN\pi$--$\pi Y\pi$ coupled channel Faddeev equations 
were solved. Table IV lists the energy eigenvalues with respect to the 
$\bar KN\pi$ threshold obtained for the models of the $\bar KN$, $\bar K\pi$, 
$\pi N$, and $\pi\pi$ interactions specified in Tables I and III, and models 
of the $I=1$ $\pi Y$ $p$-wave interaction given in Table II. Singling out 
$\pi Y$ form factors in the range $A=0.25$ to $A=0.45$ fm$^2$ from Table II, 
with values of $r_0$ up to 0.15 fm larger than the $\pi N$ value $r_0=1.36$ 
fm from Table III, as discussed in subsection \ref{subsec:size}, 
a $I(J^P)=1({\frac{3}{2}}^-)$ $\bar KN\pi$ resonance or quasibound state is 
predicted near the $\bar KN\pi$ threshold, with mass $M \approx (1570\pm 10)$ 
MeV and partial decay width $\Gamma_{\downarrow} \approx (60\pm 10)$ MeV. 
This partial width excludes contributions from channels disregarded in the 
present $\bar KN\pi$--$\pi Y\pi$ three-body model. The total width 
$\Gamma = \Gamma_{\downarrow} + \Gamma_{\uparrow}$ includes also a partial 
width $\Gamma_{\uparrow}$ induced by meson absorption modes into two-body 
$\bar K N$ and $\pi Y$ final states which were disregarded in this 
exploratory calculation and could amount to several tens of MeV.

\section{Summary} 
\label{sec:summ} 

In this work we have derived three-body Faddeev equations for the 
$\bar KN\pi$--$\pi Y\pi$ coupled channels, with isobar model separable 
$p$-wave interactions for the two-body $\pi N$ ($\Delta(1232)$), 
$\pi \bar K$ ($K^*(892)$), $\pi Y$ ($\Sigma(1385)$) and $\pi \pi$ 
($\rho(770)$) subsystems, and separable $s$-wave coupled channel 
interactions for the two-body $\bar KN$--$\pi Y$ subsystem dominated 
by the $\Lambda(1405)$ resonance. We solved these Faddeev equations, 
with two-body separable interactions fitted to available data, searching 
for a $I(J^P)=1({\frac{3}{2}}^-)$ $\bar KN\pi$ quasibound state. All the 
resonating two-body subsystems used for input in the coupled channels 
Faddeev calculation were found indispensable to obtain a quasibound 
solution near the $\bar KN\pi$ threshold. The uncertainty of this 
calculation is determined by the lack of a reliable $p$-wave phase 
shift parametrization for the $\Sigma(1385) \to \pi Y$ decay spectrum. 
Requiring a `size' of the $\pi Y$ form factor similar to that for $\pi N$, 
a $\bar KN\pi$ resonance or quasibound state exists with mass $M \approx 
(1570 \pm 10)$ MeV and decay width which is bounded from below by 
$\Gamma_{\downarrow}\approx (60\pm 10)$ MeV. With respect to two-body 
$\bar K N$ and $\pi Y$ final states, which are outside the scope of the 
present three-body model, this $\bar KN\pi$ state defines a $D_{13}$ $\Sigma$ 
resonance near the $\bar KN\pi$ threshold. The PDG listing \cite{PDG} leaves 
room for a resonance in this mass range, with the two-star `bumps' 
$\Sigma(1560)$ and the one-star $I(J^P)=1({\frac{3}{2}}^-)$ $\Sigma(1580)$ 
as possible candidates. In particular, the known two-body $\bar K N$ and 
$\pi Y$ decay branching ratios are abnormally small for $\Sigma(1580)$, 
indicating that their partial width $\Gamma_{\uparrow}$ is smaller than 
the partial width $\Gamma_{\downarrow}$ for two-meson decay modes 
$\bar KN\pi \to \pi Y\pi$ of the present model.

\begin{acknowledgments} 
This work was supported in part by the EU initiative FP7, HadronPhysics2, 
under Project No. 227431, and in part by COFAA-IPN (M\'exico). 
\end{acknowledgments}

\end{document}